%% file: dijet_search_prl_1b.tex
\documentstyle[12pt,epsf]{article}
\begin{document}
\vspace*{-0.5in}
\begin{flushright}
Fermilab-PUB-97/023-E\\
\end{flushright}
\begin{center}
\begin{Large}
{\bf  Search for New Particles Decaying to Dijets at CDF}
\end{Large}
\end{center}
\input{run1b_authors}

\renewcommand{\baselinestretch}{2}
\large
\normalsize

\clearpage

\begin{center}
{\bf Abstract}
\end{center}
We have used $106$ pb$^{-1}$ of data collected with the Collider Detector at
Fermilab to search for new particles decaying to dijets.
We exclude at the 95\% confidence level models containing the following new 
particles: 
axigluons and flavor universal colorons with mass between 200 and 980 
GeV/c$^2$, excited quarks with mass between 80 and 570~GeV/c$^2$ and between
580 and 760 GeV/c$^2$, color octet 
technirhos with mass between 260 and 480 GeV/c$^2$, $W^\prime$ bosons with
mass between 300 and 420 GeV/c$^2$, and $E_6$ diquarks with mass between
290 and 420 GeV/c$^2$.\\

\vspace*{0.5in}

In this paper we extend a previous search~\cite{ref_dijet_prl_1a} for narrow 
resonances in the dijet mass spectrum in $p\bar{p}$ collisions at a 
center-of-mass energy $\sqrt{s}=1.8$ TeV. The previous search used 
19 pb$^{-1}$ of data
collected in 1992-93 from run 1A of the Tevatron.  This search uses 106 
pb$^{-1}$ of data collected in 1992-95 from both run 1A and run 1B, and 
significantly extends our sensitivity to new particles.

As before, we perform both a general search
for narrow resonances and a specific search for axigluons~\cite{ref_axi}, 
excited quarks~\cite{ref_qstar}, color octet technirhos~\cite{ref_trho}, 
W$^{\prime}$, Z$^{\prime}$~\cite{ref_gauge}, and 
$E_6$ diquarks~\cite{ref_diquark}. In addition, the flavor universal 
coloron~\cite{ref_coloron}, a hypothesized massive gluon which couples equally
to all quarks, is considered together with axigluons. The cross section for 
the coloron is always greater than or equal to that of 
the axigluon, so our axigluon limits will apply to the coloron as well.
In models of supersymmetry in which the gluino is lighter than 5 GeV/c$^2$, 
there can be dijet resonances resulting from squark 
decay~\cite{ref_clavelli, ref_hewett}. We do not consider this model, since 
data from both our previous search and from a 
preliminary version of the present search has already been used to exclude 
a range of squark masses in the light gluino 
scenario~\cite{ref_clavelli, ref_hewett}.

A detailed description of the Collider Detector at Fermilab (CDF) can be found 
elsewhere~\cite{ref_CDF}. We use a coordinate system with the $z$ axis along the proton
beam, transverse coordinate perpendicular to the beam, azimuthal angle $\phi$, 
polar angle $\theta$, and pseudorapidity $\eta=-\ln \tan(\theta/2)$. 
Jets are reconstructed as localized energy depositions in the CDF calorimeters
that are arranged in a projective tower geometry.
The jet energy $E$ is defined as the scalar sum of the calorimeter tower 
energies inside a cone of radius 
$R=\sqrt{(\Delta\eta)^2 + (\Delta\phi)^2}=0.7$, centered on the jet direction. 
The jet momentum $\vec{P}$ is the corresponding vector sum: 
$\vec{P} = \sum{E_i\hat{u}_i}$ with $\hat{u}_i$ being the 
unit vector pointing from the interaction point to the energy deposition $E_i$ 
inside the same cone. $E$ and $\vec{P}$ are corrected 
for calorimeter non-linearities, energy lost in uninstrumented regions of the
detector and outside the clustering cone, and
energy gained from the underlying event and multiple $p\bar{p}$ interactions. 
The jet energy corrections increase the jet energies on average by roughly
24\% (19\%) for 50 GeV (500 GeV) jets.  Full details of jet reconstruction and
jet energy corrections at CDF can be found elsewhere~\cite{ref_jet}.

We define the dijet system as the
two jets with the highest transverse momentum in an event (leading jets) 
and define the dijet mass
$m=\sqrt{(E_1 + E_2)^2 - (\vec{P}_1 + \vec{P}_2)^2}$.  The dijet mass
resolution is approximately 10\% for dijet mass above 150 GeV/c$^2$.
Our data sample was obtained using four triggers
that required at least one jet with uncorrected cluster transverse energies
of 20, 50, 70 and 100 GeV, respectively.  After jet energy corrections these 
trigger samples
were used to measure the dijet mass spectrum above 180, 241, 292 and 388 
GeV/c$^2$, respectively.  At these mass thresholds the trigger efficiencies were 
greater than 95\%. The four data samples 
corresponded to integrated luminosities of $0.126$, $2.84$, $14.1$ and $106$ 
pb$^{-1}$ after prescaling.  Offline we required that
both jets have pseudorapidity $|\eta|<2$ and a scattering angle in the dijet 
center-of-mass frame 
$|\cos\theta^*| = |\tanh[(\eta_1-\eta_2)/2]| < 2/3$. The $\cos\theta^*$ 
requirement provides uniform acceptance as a function of mass and reduces the 
QCD background which  peaks at $|\cos\theta^*|=1$.  
To utilize the projective nature of the calorimeter towers, the $z$ position
of the event vertex was required to be within 60 cm of the center of the
detector; this cut removed 7\% of the events. Backgrounds from cosmic-rays, 
beam halo, and detector noise were removed with the cuts reported 
previously~\cite{ref_dijet_prl_1a}, and residual backgrounds were removed by 
requiring that the total observed energy be less than 2 TeV.                   

\begin{figure}[tbh]
\hspace*{-0.5in}
\vspace*{-1.2in}
\epsfysize=7.5in
\epsffile[36 61 540 650]{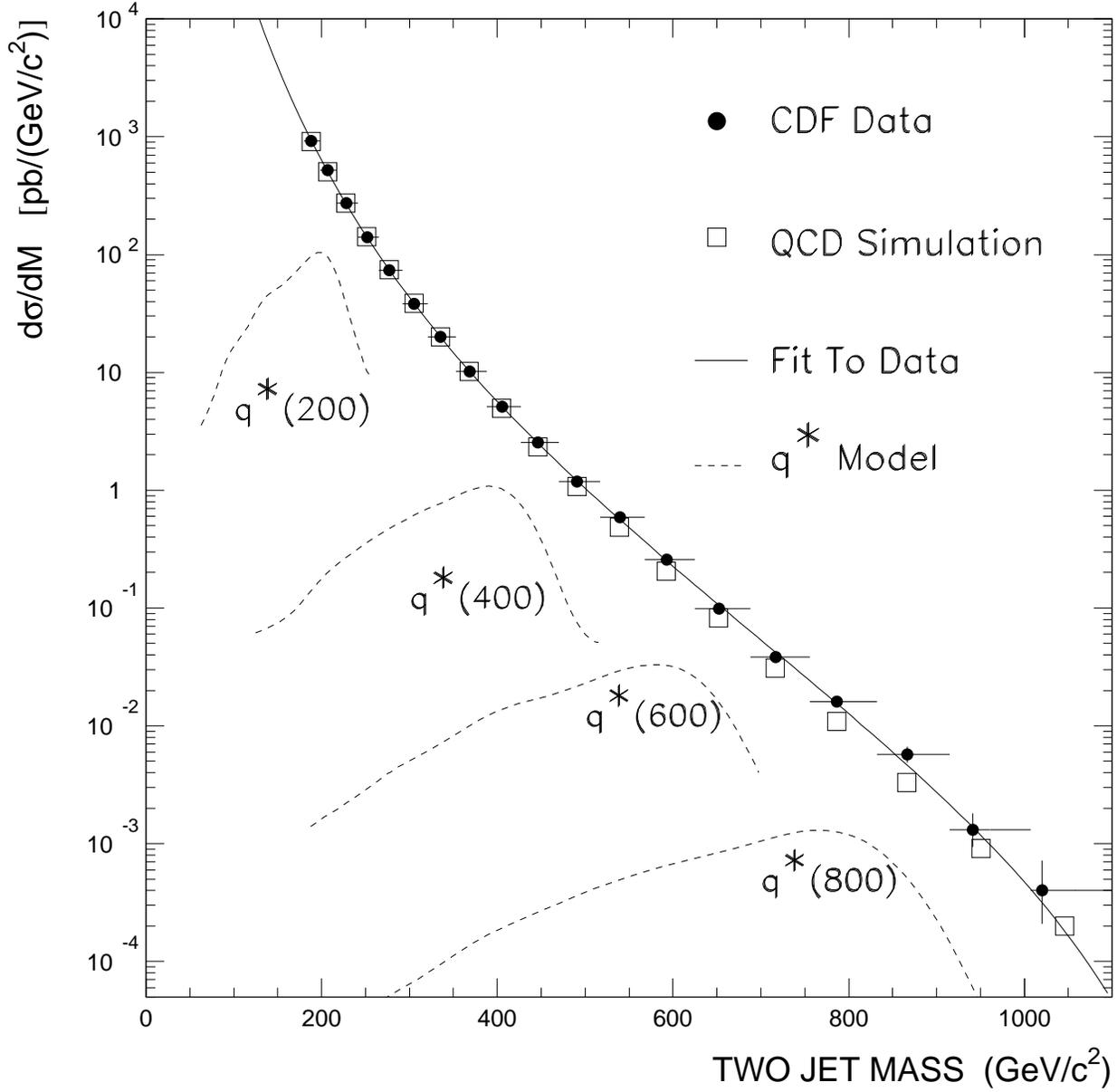}
\caption[Dijet Mass, Background and Excited Quark Signal]{ 
The dijet mass distribution (circles) compared to a QCD simulation (boxes) and 
fit to a smooth parameterization (solid curve). Also shown are simulations 
of excited quark signals in the 
CDF detector (dashed curves). In the data and simulations we require that both 
jets have pseudorapidity $|\eta|<2.0$ and that the dijet system satisfies 
$|\cos\theta^*|<2/3$.}
\label{fig_mass_qstar}
\end{figure}

In Fig.~\ref{fig_mass_qstar} we present the inclusive dijet mass 
distribution for $p\bar{p}\rightarrow$ 2 jets + X, where X can be 
anything including additional jets. 
The dijet mass
distribution has been corrected for trigger and $z$ vertex inefficiencies.
We plot the differential cross section versus the mean dijet mass in 
bins of width approximately equal to the dijet mass resolution (RMS$\sim 10$\%). 
The data are compared to a QCD prediction from the PYTHIA 
Monte Carlo~\cite{ref_pythia} and a simulation of the CDF detector.
The cross section predicted by the QCD simulation, using CTEQ2L parton 
distributions~\cite{ref_CTEQ} and a renormalization scale $\mu=P_T$, 
is normalized to the data in the first 6 bins ($180<m<321$ GeV/c$^2$) by 
dividing the simulation by a factor of $0.66$. In Fig.~\ref{fig_mass_qstar} 
the horizontal lines on the data points indicate the bin width, the same width 
in data and simulation. The points are plotted at the mean mass,
calculated independently for data and simulation.

We note that the data is above the QCD simulation at high mass. 
In a previous paper~\cite{ref_jet_prl_1a}, 
we reported a similar effect in the fully corrected inclusive jet 
transverse energy distribution compared to an O$(\alpha_s^3)$ parton level QCD 
calculation.  Unlike the inclusive jet analysis, here we do not deconvolute 
the mass distribution for the effects of detector resolution, and instead 
compare the data directly to QCD plus a CDF detector simulation. In our 
previous dijet mass search the excess was not as noticeable because we 
normalized the simulation to the data on average, while here we normalize to
the low mass end as described above.  In another paper~\cite{ref_ang_dist} we 
have studied the
dijet angular distributions and find them to be in good agreement with QCD
in all regions, including at high mass.  The source of the high dijet mass
and high jet transverse energy excess is not yet fully understood.
Candidate explanations within the Standard Model include 
a larger than expected gluon distribution of the 
proton~\cite{ref_new_gluon} or large QCD corrections from 
resummation~\cite{ref_qcd_corr}.
As in our previous search~\cite{ref_dijet_prl_1a}, we do not use QCD 
calculations to determine 
the background to new particles, but merely use the data itself to fit for
the background.

\begin{figure}[tbh]
\hspace*{-0.5in}
\vspace*{-1.0in}
\epsfysize=7.5in
\epsffile[36 61 540 650]{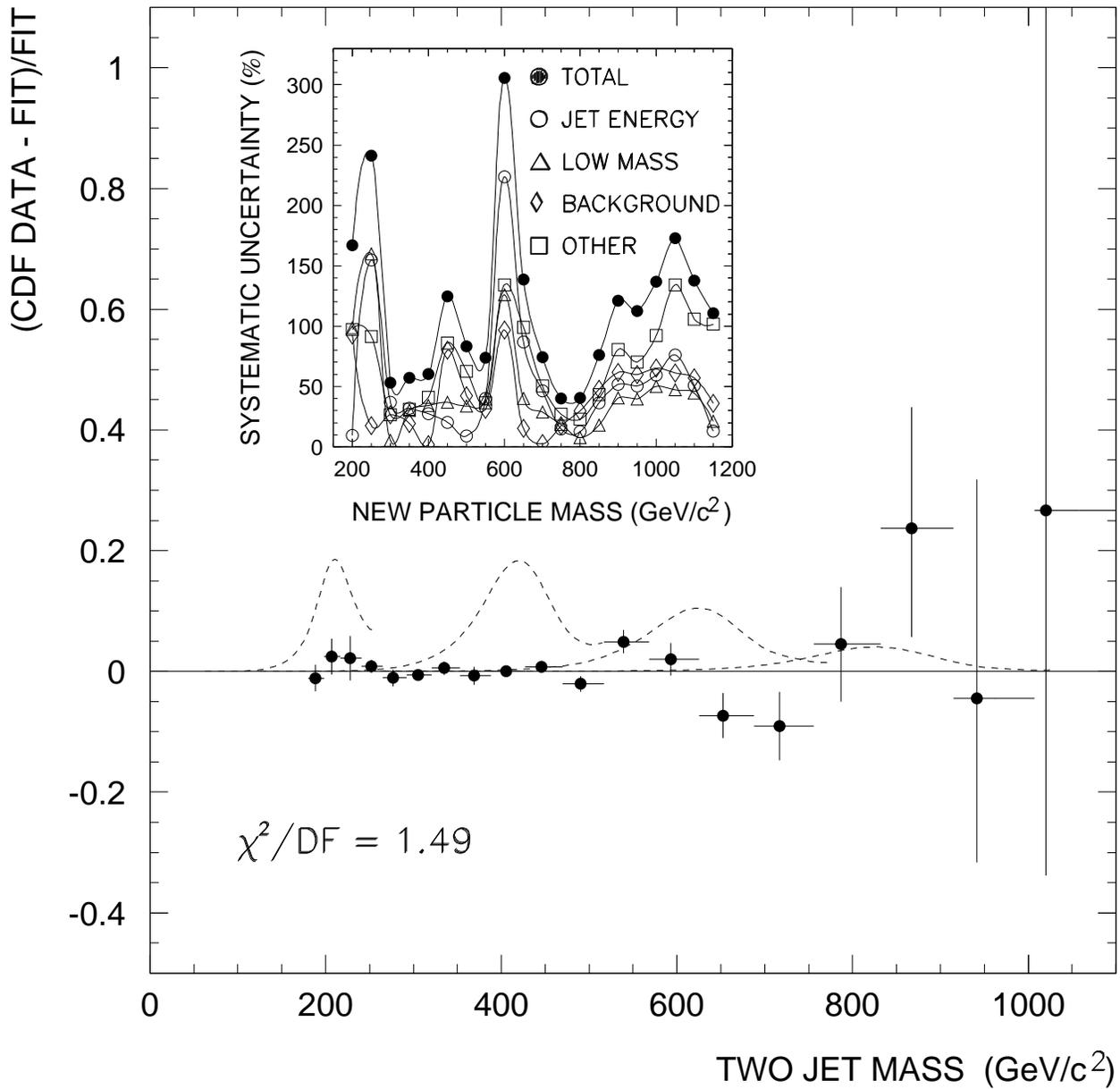}
\caption[Dijet Mass and Background Fit on a Linear Scale]{ 
The fractional difference between the dijet mass distribution (points) and
a smooth background fit (solid line) is compared to simulations of 
excited quark signals in the CDF detector (dashed curves). The inset shows the 
systematic uncertainty for a new particle signal (see text).}
\label{fig_mass_lin}
\end{figure}

To search for resonances we fit the data with the parameterization 
$d\sigma/dm = A(1-m/\sqrt{s}+Cm^2/s)^N/m^P$ with parameters  A, C, N and P.  
In the run 1A search~\cite{ref_dijet_prl_1a} the term $Cm^2/s$ was not used 
because fewer parameters were needed to fit the lower statistics sample.
With the higher statistics in this sample the extra term
$Cm^2/s$ was needed to obtain an acceptable fit.
This parameterization gives an adequate description of both the observed
distribution ($\chi^2/DF=1.49$) and the QCD prediction ($\chi^2/$DF$=0.85$). 
Figure ~\ref{fig_mass_qstar} shows the background fit on a 
logarithmic scale, and Fig.~\ref{fig_mass_lin} shows the fractional difference 
between the data and background fit on a linear scale.  

Figures ~\ref{fig_mass_qstar} and ~\ref{fig_mass_lin} also show the predicted 
line shape for excited quarks ($q^*$) using 
the PYTHIA Monte Carlo~\cite{ref_pythia} and a CDF detector simulation. 
If excited quarks were produced in $p\bar{p}$ collisions, their production and
decay to dijets would proceed via the process 
$qg \rightarrow q^* \rightarrow qg$. The mass resolution is 
dominated by a Gaussian distribution (RMS$\sim 10$\%) 
from jet energy resolution and a long tail towards low mass from QCD radiation. 
Since the natural width of a $q^*$ is
significantly smaller than the measured width, the $q^*$ mass resonance curves in 
Figs.~\ref{fig_mass_qstar} and 
\ref{fig_mass_lin} were used to model the shape of all narrow resonances 
decaying to dijets. 

There is no statistically significant evidence for a dijet
mass resonance, which should appear in at least two neighboring bins above 
the background fit. We note that in the region of 550 GeV/c$^2$ there is a 
single bin 
which is $2.6$ standard deviations above the fit; however, this region is not 
well fit by a new resonance because the number of events in neighboring bins is 
too low.  When we fit the data to both a 550 GeV/c$^2$ resonance and a smooth
background we find that the upward fluctuation in the data is significantly 
narrower than expected for a resonance.  

Systematic uncertainties on the cross section for observing a new particle 
in the CDF detector are shown in Fig.~\ref{fig_mass_lin}.  
Each systematic uncertainty on the fitted signal cross section was determined 
by varying the source of uncertainty by $\pm 1\sigma$ and refitting.
In decreasing order of importance the sources 
of uncertainty are the 5\% jet energy scale 
uncertainty, low mass data, the background parameterization,
QCD radiation's effect on the mass resonance line 
shape, trigger efficiency, jet energy
resolution, relative jet energy corrections between different parts of the 
CDF calorimeter, energy scale of run 1A with respect to run 1B, luminosity and 
efficiency. 
For example, at 600 GeV/c$^2$ reducing the jet energy by 5\% centers the 
resonance on an upward fluctuation, and increases the fitted signal by 225\%.
The low mass data uncertainty, listed above, is because the
background fit gets significantly worse when data between 150 and 180 GeV/c$^2$
are included. The larger number
of interactions per crossing in run 1B increases the uncertainty on the lower 
mass data, so we start the mass distribution at 180 GeV/c$^2$.
However, since this mass range was included in run 1A, the
effect of adding the low mass data is included as a systematic for run 1A plus
run 1B. 

The total systematic uncertainty was found by adding the individual 
sources in quadrature. 
In this analysis the relative systematic error is larger than it was in the 
previous analysis: the total run 1A 
and 1B systematics range from 40\% to 300\% of the 
cross section while the run 1A systematics ranged from 30\% to 120\%.
This is not because the absolute systematics have significantly 
increased, but instead because the size of the signal we are statistically 
sensitive to has decreased by over a factor of two, so now the systematics
have a larger relative effect. This is particularly true at masses near 
upward fluctuations in the data.  

In the absence of conclusive evidence for new physics we proceed to set upper
limits on the cross section for new particles. 
For each value of new particle mass in 50 GeV/c$^2$ steps from 200 to 1150
GeV/c$^2$, we perform a binned maximum likelihood fit of the data to the 
background parameterization and the mass resonance shape. 
We convolute each of the 20
likelihood distributions with the corresponding total Gaussian systematic 
uncertainty, and find the 95\% confidence level 
(CL) upper limit presented in Table I.

\begin{figure}[tbh]
\hspace*{-0.5in}
\vspace*{-1.3in}
\epsfysize=7.5in
\epsffile[36 61 540 650]{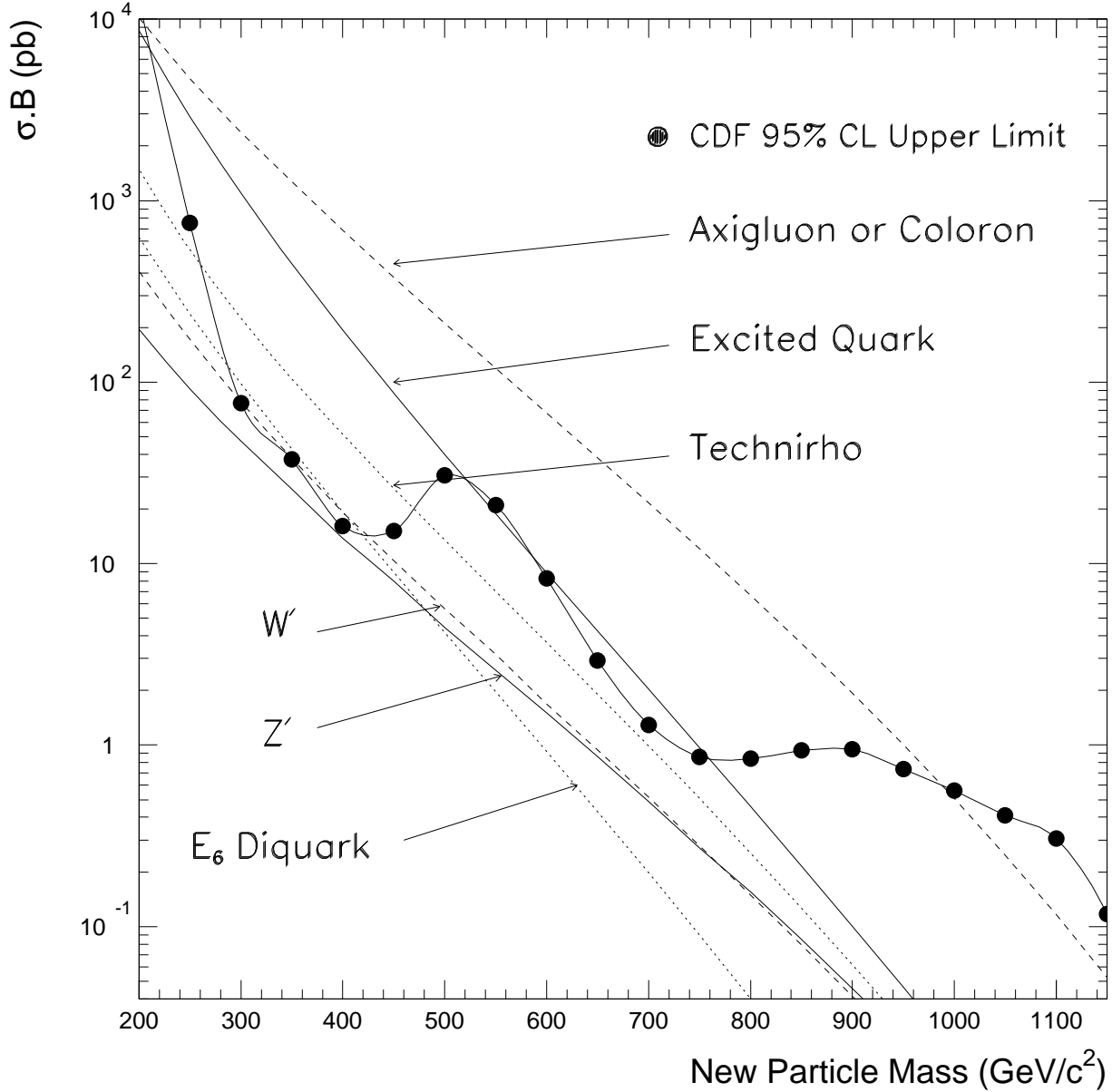}
\caption[Upper Limit on Cross Section for New Particles]{ 
The upper limit on the cross section times branching 
ratio for new particles decaying to dijets (points) is compared to 
theoretical predictions for axigluons~\cite{ref_axi}, flavor universal
colorons~\cite{ref_coloron}, excited
quarks~\cite{ref_qstar}, color octet technirhos~\cite{ref_trho}, new
gauge bosons $W^{\prime}$ and $Z^{\prime}$~\cite{ref_gauge}, and 
$E_6$ diquarks~\cite{ref_diquark}. The limit and theory curves require that both 
jets have pseudorapidity $|\eta|<2.0$ and that the dijet system satisfies 
$|\cos\theta^*|<2/3$.}                                           
\label{fig_limit}
\end{figure}

\begin{figure}[tbh]
\hspace*{-0.1in}
\vspace*{-0.2in}
\epsfysize=6.6in
\epsffile[72 144 550 650]{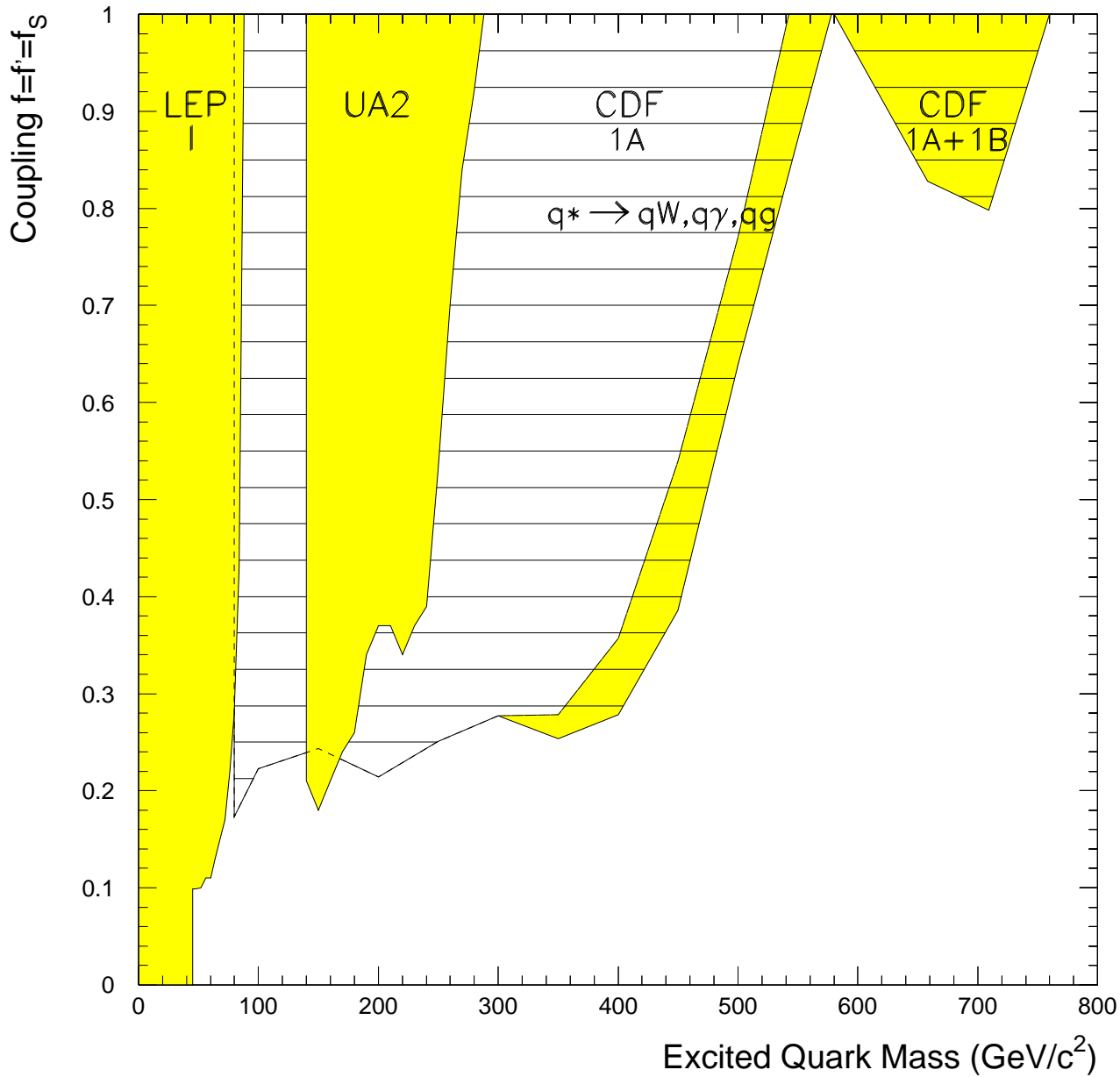}
\caption[Limit on Excited Quarks in Mass vs. Coupling]{ 
The region of the coupling vs. mass plane excluded by previous CDF 
measurements~\cite{ref_dijet_prl_1a,ref_qstar_prl} in the $q^*\rightarrow 
q\gamma$ and $q^*\rightarrow qW$
channels (clear hatched region) and
$q^*\rightarrow qg$ channels (shaded hatched region on left) in run 1A is 
extended by this $q^*\rightarrow qg$ search in run 1A plus run 1B (shaded 
hatched region on right). The CDF excluded regions 
are compared to the regions excluded by LEP I and UA2 
(shaded regions)~\cite{ref_other}.}
\label{fig_coupling}
\end{figure}

In Fig.~\ref{fig_limit} we plot our measured upper limit on the cross section 
times branching ratio for a new particle decaying to dijets as a function of
new particle mass in 50 GeV/c$^2$ steps.  The points are connected by a smooth
curve, which is an estimate of the upper limit in between the measured points.
The limit is compared to lowest order theoretical predictions for the cross 
section times branching ratio for new particles decaying to 
dijets~\cite{ref_dijet_prl_1a}.
New particle decay angular distributions are included in the calculations, and 
we  required $|\eta|<2$ and $|\cos\theta^*|<2/3$ for all predictions. 
For axigluons (or flavor universal colorons) 
we exclude the mass range $200<M_A<980$ GeV/c$^2$, extending the
previous CDF exclusions of $120<M_A<870$
GeV/c$^2$~\cite{ref_dijet_prl_1a}.
For excited quarks we exclude the mass
ranges $200<M^*<520$ and $580<M^*<760$ GeV/c$^2$, significantly extending the 
previous CDF exclusion of
$80<M^*<570$ GeV/c$^2$~\cite{ref_dijet_prl_1a,ref_qstar_prl}. 
The D0 collaboration has performed a preliminary search for excited quarks and 
exclude the mass range $200<M^*<720$ GeV/c$^2$~\cite{ref_D0}.
These exclusions are for Standard Model couplings ($f=f^\prime=f_s=1$). For
smaller couplings, the 
new excluded region in the coupling~\cite{ref_qstar} vs. mass plane 
is shown in Fig.~\ref{fig_coupling} compared to previous excluded regions. 
For color octet technirhos ($\rho_T$) we exclude the mass range 
$260<M_{\rho_T}<470$ GeV/c$^2$, extending to lower mass the previous CDF 
exclusion of
$320<M_{\rho_T}<480$ GeV/c$^2$~\cite{ref_dijet_prl_1a}. 
For the first time we exclude the hadronic decays
of the new gauge boson W$^\prime$
in the mass range $300<M_{W^\prime}<420$ GeV/c$^2$. Also for the first time
we exclude $E_6$ diquarks in the mass range $290<M_{E_6}< 420$ GeV/c$^2$. 
The cross section for hadronic decays of $Z^{\prime}$ is too small to exclude.

In conclusion, the measured dijet mass spectrum does not contain 
evidence for a mass peak from a new particle resonance. We have 
presented model independent limits on the cross section for a narrow resonance, 
and set specific mass 
limits on axigluons, flavor universal colorons, excited quarks, color octet 
technirhos, new charged gauge bosons, and $E_6$ diquarks.

We thank the Fermilab staff and the technical staffs of the participating 
institutions for their vital contributions. This work was
supported by the U.S. Department of Energy and National Science Foundation;
the Italian Istituto Nazionale di Fisica Nucleare; the Ministry of Education, 
Science and Culture of Japan; the Natural Sciences and Engineering Research
Council of Canada; the National Science Council of the Republic of China; 
and the A. P. Sloan Foundation.

\clearpage

\renewcommand{\baselinestretch}{1.4}
\large
\normalsize

\begin{table}[tbh]
\begin{center}
\begin{tabular}{|c|c|c|l|}\hline 
Mass   &  95\% CL & Mass  & 95\% CL \\
 (GeV/c$^2$) & $\sigma\cdot B$ (pb) &   (GeV/c$^2$)  & $\sigma\cdot B$ (pb) \\ \hline
 200 & $1.3\times 10^4$   & 700 & $1.3\times 10^0$ \\
 250 & $7.6\times 10^2$   & 750 & $8.6\times 10^{-1}$ \\
 300 & $7.7\times 10^1$   & 800 & $8.4\times 10^{-1}$ \\
 350 & $3.8\times 10^1$   & 850 & $9.3\times 10^{-1}$ \\
 400 & $1.6\times 10^1$   & 900 & $9.5\times 10^{-1}$ \\
 450 & $1.5\times 10^1$   & 950 & $7.4\times 10^{-1}$ \\
 500 & $3.1\times 10^1$   & 1000 & $5.6\times 10^{-1}$ \\
 550 & $2.1\times 10^1$   & 1050 & $4.1\times 10^{-1}$ \\
 600 & $8.3\times 10^0$   & 1100 & $3.1\times 10^{-1}$ \\
 650 & $2.9\times 10^0$   & 1150 & $1.2\times 10^{-1}$ \\
\hline
\end{tabular}
\end{center}
\label{tab_limit}
\end{table}

\renewcommand{\baselinestretch}{2}
\large
\normalsize
\vspace*{-0.3in}
Table I: As a function of new particle mass we list our 95\% CL upper limit on 
cross section times branching ratio for narrow resonances decaying to dijets.
The limit applies to the kinematic range where both jets have pseudorapidity 
$|\eta|<2.0$ and where the dijet system satisfies $|\cos\theta^*|<2/3$.
\clearpage

\clearpage

\clearpage

\clearpage

\end{document}

%% file: run1b_authors.tex
\font\eightit=cmti8
\def\r#1{\ignorespaces $^{#1}$}
\hfilneg
\begin{sloppypar}
\noindent
F.~Abe,\r {16} H.~Akimoto,\r {35}
A.~Akopian,\r {30} M.~G.~Albrow,\r 7 S.~R.~Amendolia,\r {26} 
D.~Amidei,\r {19} J.~Antos,\r {32} S.~Aota,\r {35}
G.~Apollinari,\r {30} T.~Asakawa,\r {35}
% ...
\end{sloppypar}

\vskip .026in
\begin{center}
(CDF Collaboration)
\end{center}

\vskip .026in
\begin{center}
\r 1  {\eightit Argonne National Laboratory, Argonne, Illinois 60439} \\
\r 2  {\eightit Istituto Nazionale di Fisica Nucleare, University of Bologna,
I-40127 Bologna, Italy} \\
\r 3  {\eightit Brandeis University, Waltham, Massachusetts 02264} \\
\r 4  {\eightit University of California at Los Angeles, Los 
Angeles, California  90024} \\  
\r 5  {\eightit University of Chicago, Chicago, Illinois 60638} \\
\r 6  {\eightit Duke University, Durham, North Carolina  28708} \\
\r 7  {\eightit Fermi National Accelerator Laboratory, Batavia, Illinois 
60510} \\
\r 8  {\eightit University of Florida, Gainesville, FL  33611} \\
\r 9  {\eightit Laboratori Nazionali di Frascati, Istituto Nazionale di Fisica
               Nucleare, I-00044 Frascati, Italy} \\
\r {10} {\eightit University of Geneva, CH-1211 Geneva 4, Switzerland} \\
\r {11} {\eightit Harvard University, Cambridge, Massachusetts 02138} \\
\r {12} {\eightit Hiroshima University, Higashi-Hiroshima 724, Japan} \\
\r {13} {\eightit University of Illinois, Urbana, Illinois 61801} \\
\r {14} {\eightit Institute of Particle Physics, McGill University, Montreal 
H3A 2T8, and University of Toronto,\\ Toronto M5S 1A7, Canada} \\
\r {15} {\eightit The Johns Hopkins University, Baltimore, Maryland 21218} \\
\r {16} {\eightit National Laboratory for High Energy Physics (KEK), Tsukuba, 
Ibaraki 315, Japan} \\
\r {17} {\eightit Ernest Orlando Lawrence Berkeley National Laboratory, 
Berkeley, California 94720} \\
\r {18} {\eightit Massachusetts Institute of Technology, Cambridge,
Massachusetts  02139} \\   
\r {19} {\eightit University of Michigan, Ann Arbor, Michigan 48109} \\
\r {20} {\eightit Michigan State University, East Lansing, Michigan  48824} \\
\r {21} {\eightit University of New Mexico, Albuquerque, New Mexico 87132} \\
\r {22} {\eightit The Ohio State University, Columbus, OH 43320} \\
\r {23} {\eightit Osaka City University, Osaka 588, Japan} \\
\r {24} {\eightit Universita di Padova, Istituto Nazionale di Fisica 
          Nucleare, Sezione di Padova, I-36132 Padova, Italy} \\
\r {25} {\eightit University of Pennsylvania, Philadelphia, 
        Pennsylvania 19104} \\   
\r {26} {\eightit Istituto Nazionale di Fisica Nucleare, University and Scuola
               Normale Superiore of Pisa, I-56100 Pisa, Italy} \\
\r {27} {\eightit University of Pittsburgh, Pittsburgh, Pennsylvania 15270} \\
\r {28} {\eightit Purdue University, West Lafayette, Indiana 47907} \\
\r {29} {\eightit University of Rochester, Rochester, New York 14628} \\
\r {30} {\eightit Rockefeller University, New York, New York 10021} \\
\r {31} {\eightit Rutgers University, Piscataway, New Jersey 08854} \\
\r {32} {\eightit Academia Sinica, Taipei, Taiwan 11530, Republic of China} \\
\r {33} {\eightit Texas A\&M University, College Station, Texas 77843} \\
\r {34} {\eightit Texas Tech University, Lubbock, Texas 79409} \\
\r {35} {\eightit University of Tsukuba, Tsukuba, Ibaraki 315, Japan} \\
\r {36} {\eightit Tufts University, Medford, Massachusetts 02155} \\
\r {37} {\eightit University of Wisconsin, Madison, Wisconsin 53806} \\
\r {38} {\eightit Yale University, New Haven, Connecticut 06511} \\
\end{center}

%% file: dijet_search_prl_1b.bbl
\begin{thebibliography}{99.}

\bibitem{ref_dijet_prl_1a} F. Abe {\em et al.}, Phys. Rev. Lett. {\bf 74}, 3538 
(1995).

\bibitem{ref_axi} P. Frampton and S. Glashow, Phys. Lett. {\bf B190}, 157 (1987);
J. Bagger, C. Schmidt and S. King, Phys. Rev. {\bf D37}, 1188 (1988).

\bibitem{ref_qstar} U. Baur, I. Hinchliffe and D. Zeppenfeld, Int. J. Mod. 
Phys {\bf A2}, 1285 (1987); U. Baur, M. Spira and P. Zerwas, Phys. Rev. {\bf 
D42}, 815 (1990).

\bibitem{ref_trho} K. Lane and M. Ramana, Phys. Rev. {\bf D44}, 2678 (1991); 
E. Eichten and K. Lane, Phys. Lett. {\bf B327}, 129 (1994).

\bibitem{ref_gauge} F. Abe {\em et al.}, Phys. Rev. Lett. {\bf 74}, 2900 (1995) and
Phys. Rev. {\bf D51}, R949 (1995), and references therein.

\bibitem{ref_diquark} J. Hewett and T. Rizzo, 
Phys. Rep. {\bf 183}, 193 (1989) and references therein.

\bibitem{ref_coloron} R. S. Chivukula, A. G. Cohen, and E. Simmons, Phys. Lett. 
{\bf B380}, 92 (1996); E. Simmons, Phys. Rev. {\bf D55}, 1678 (1997).

\bibitem{ref_clavelli} I. Terekhov and L. Clavelli, Phys. Lett. {\bf B385}, 139
(1996).

\bibitem{ref_hewett} J. Hewett, T. Rizzo and M. Doncheski, SLAC-PUB-7372 (1996).

\bibitem{ref_CDF} F.\ Abe {\em et al.}, Nucl.\ Instrum.\ and Methods {\bf A271}, 387 (1988).

\bibitem{ref_jet} F. Abe {\em et al.}, Phys. Rev. {\bf D45}, 1448 (1992).

\bibitem{ref_pythia} PYTHIA $V5.6$ by T.\ Sjostrand, CERN-TH-7112/93, Feb 1994.

\bibitem{ref_CTEQ}  J. Botts {\em et al.}, Phys. Lett. {\bf B304}, 159 (1993).

\bibitem{ref_jet_prl_1a} F.\ Abe {\em et al.}, Phys. Rev. Lett. {\bf 77}, 438 (1996).

\bibitem{ref_ang_dist} F.\ Abe {\em et al.}, Phys. Rev. Lett. {\bf 77}, 5336 (1996).

\bibitem{ref_new_gluon} H. L. Lai {\em et al.}, MSUHEP-60426, hep-ph/9606399
submitted to Phys. Rev. D.

\bibitem{ref_qcd_corr} S. Catani {\em et al.}, Nucl. Phys. {\bf B478}, 273 (1996). 

\bibitem{ref_qstar_prl} F. Abe {\em et al.}, Phys. Rev. Lett. {\bf 72}, 3004 (1994).

\bibitem{ref_D0} I. Bertram for the D0 Collaboration, Fermilab-Conf-96/389-E 
(1996).

\bibitem{ref_other} D.\ Decamp {\em et al.}, Phys. Rep. {\bf 216}, 253 (1992); 
J.\ Alitti {\em et al.}, Nucl. Phys. {\bf B400}, 3 (1993).

\end{thebibliography}
